\documentclass[fleqn,usenatbib]{mnras}
 \pdfoutput=1
\usepackage{newtxtext,newtxmath,hyperref}
\usepackage[T1]{fontenc}
\usepackage{graphicx}	
\usepackage{amsmath}	
\usepackage{amssymb}	
\usepackage[english]{babel}
\usepackage{gensymb}

\title[Polarized contamination of the global 21~cm signal]{On the contamination of the global 21~cm signal from polarized foregrounds}

\author[M. Spinelli et al.]{Marta~Spinelli$^{1,2}$\thanks{E-mail: marta.spinelli@inaf.it}, Gianni~Bernardi$^{3,4,5}$ and Mario~G.~Santos$^{2,5}$ 
\\
$^{1}$INAF-Osservatorio Astronomico di Trieste, Via G.B. Tiepolo 11, 34143 Trieste, Italy\\
$^{2}$Department of Physics and Astronomy, University of Western Cape, Cape Town 7535, South Africa\\
$^{3}$INAF-Istituto di Radioastronomia, via Gobetti 101, 40129, Bologna, Italy\\
$^{4}$Department of Physics and Electronics, Rhodes University, PO Box 94, Grahamstown, 6140, South Africa\\
$^{5}$South African Radio Astronomy Observatory, Black River Park, 2 Fir Street, Observatory, Cape Town, 7925, South Africa
}
\date{Accepted XXX. Received YYY; in original form ZZZ}

\pubyear{2019}
\begin{document}
\label{firstpage}
\pagerange{\pageref{firstpage}--\pageref{lastpage}}
\maketitle

\begin{abstract}
Global (i.e. sky-averaged) $21$~cm signal experiments can measure the evolution of the universe from the Cosmic Dawn to the Epoch of Reionization. These measurements are challenged by the presence of bright foreground emission that can be separated from the cosmological signal if its spectrum is smooth. This assumption fails in the case of single polarization antennas as they measure linearly polarized foreground emission - which is inevitably Faraday rotated through the interstellar medium.
We investigate the impact of Galactic polarized foregrounds on the extraction of the global 21~cm signal through realistic sky and dipole simulations both in a low frequency band from $50$ to $100$~MHz, where a 21~cm absorption profile is expected, and in a higher frequency band ($100-200$~MHz).
We find that the presence of a polarized contaminant with complex frequency structure can bias the amplitude and the shape of the reconstructed signal parameters in both bands. 
We investigate if polarized foregrounds can explain the unexpected $21$~cm Cosmic Dawn signal recently reported by the EDGES collaboration. 
We find that unaccounted polarized foreground contamination can produce an enhanced and distorted $21$~cm absorption trough similar to the anomalous profile reported by Bowman et al. (2018), and whose amplitude is in mild tension with the assumed input Gaussian profile (at $\sim 1.5 \sigma$ level). 
Moreover, we note that, under the hypothesis of contamination from polarized foreground, the amplitude of the reconstructed EDGES signal can be overestimated by around $30\%$, mitigating the requirement for an explanation based on exotic physics.
\end{abstract}



\section{Introduction} \label{sec:intro}
The $21$~cm background arising from the spin-flip transition  of neutral Hydrogen in the intergalactic medium is considered the most promising observable for the Cosmic Dawn and the subsequent Epoch of Reionization 
\citep[EoR; e.g.,][]{Pritchard2010}.
The $21$~cm signal is observable as a contrast
against the Cosmic Microwave Background (CMB) temperature 
\citep{Furlanetto2006}. 
As soon as the first galaxies begin to appear, they produce Ly-$\alpha$ photons that couple the excitation temperature of the $21$~cm line (spin temperature) 
to gas kinetic temperature 
through the Wouthuysen-Field effect \citep[WF,][]{Wouthuysen1952, Field1958}. 
As the gravitational collapse progresses,
the spin temperature becomes eventually completely coupled to the gas temperature 
and driven well above the CMB temperature as consequence of the gas heating - most likely by an X-ray background \citep[e.g.,][]{Venkatesan2001,Pritchard2007,Mesinger2013}.
Observations of $21$~cm {\it fluctuations} from this era of interplay between the Ly-$\alpha$ coupling and the X-ray heating will require sensitivities only achievable with the Hydrogen Epoch of Reionization Array \citep{DeBoer2017} and the upcoming Square Kilometre Array \citep{Koopmans2015}.
The measurement of the {\it global - i.e. sky averaged -} $21$~cm signal can be, conversely, achieved by a single dipole antenna observing for a few tens to a few hundreds of hours \citep[e.g.,][]{Shaver1999,Bernardi2015,Harker2016}.
The $21$~cm global signal at the Cosmic Dawn is expected to be a few hundred $\mathrm{mK}$ absorption trough depending on the offset between the WF coupling and the X-ray heating epochs \citep{Pritchard2010}. 
It is sensitive to the formation of the first luminous structures in the universe \citep[e.g.,][]{Furlanetto2006,Mirocha2014,Mesinger2016}, as well as the thermal history of the intergalactic medium \citep{Pritchard2007,Mesinger2013}.

At $z\lesssim 10-15$, the sustained galaxy formation produces an ultraviolet radiation background that eventually extinguishes  the neutral Hydrogen, and therefore the $21$~cm signal. 
The $21$~cm signal therefore traces the evolution of the average neutral fraction, essentially timing cosmic reionization.

The Experiment to Detect the Global Epoch-of-Reionization Signatures (EDGES) team has recently reported the detection of a $21$~cm absorption profile, centered at $78$~MHz, with a $19$~MHz width and an amplitude of $520$~mK \citep{EDGES}.
This result is more than a factor two stronger than standard theoretical predictions and has triggered exotic explanations like interaction with dark matter \citep[e.g.,][]{Barkana2018,Fraser2018} or Axion-Induced Cooling \citep[e.g.,][]{Houston2018} and a debate on a possible low-frequency excess radio background \citep[e.g.,][]{Ewall-Wice2018,Feng2018,Sharma2018}. 
The unexpected EDGES result is awaiting for independent confirmation from the other ongoing global signal experiments. 
These experiments include the Large aperture Experiment to detect the Dark Ages (LEDA; \citet{LEDA}) that constrained at $95\%$ level the amplitude ($>-890$~mK) and the $1\sigma$ width ($>6.5$~MHz) for a Gaussian model for the trough \citep{Bernardi2016}; the ``Sonda  Cosmologica  de  las  Islas  para  la Deteccion de Hidrogeno Neutro  \citep[SCI-HI;][]{Voytek2014} that reported a $1$~K rms residual in the range $60-88$~MHz; the upgraded  Shaped Antenna measurement of the background RAdio Spectrum (SARAS~3) that has already provided constraints in the $6 < z < 10$ range \citep{Singh2017,Singh2018}, the  Probing Radio Intensity at high-Z from Marion (PRIZM) experiment \citep{Philip2019}, and the future Dark Ages Radio Explorer \citep[DARE;][]{Mirocha2015} is planning to measure the $21$~cm global signal. This would also allow to avoid not only terrestrial radio frequency interference, but also ionospheric corruption and solar radio emissions.

The key challenge to measure the 21~cm signal is the subtraction of the bright foreground emission and the consequent control of systematic effects. In presence of smooth-spectrum foregrounds, simulations show that the 21~cm signal can generally be extracted \citep{Nhan2017,Rao2017,Singh2017,Singh2018,Tauscher2018}, particularly using Bayesian techniques \citep[e.g.,][]{Harker2012,Bernardi2015,Bernardi2016,Monsalve2017,Monsalve2018,Monsalve2019}. This strategy has been employed by \citet{EDGES} too, although their unusual findings have drawn the attention to their foreground modelling and separation method.
\citet{Hills2018} have, for example, re-examined the EDGES data and questioned their detection pointing out that the extracted foreground model parameters are unphysical. The re-analysis by \citet{Singh2019}, enforcing a maximally smooth foreground model, also found evidence for a different $21$~cm signal, substantially more in agreement with the standard predictions.

In this work, we investigate the effect that Galactic polarized foreground emission has on the measurement of the $21$~cm signal. Polarized foreground that are Faraday rotated through the interstellar medium can leak into total intensity because of imperfect calibration and can, therefore, violate the assumption of smooth spectrum foregrounds. This effect is an active subject of study for interferometric observations \citep[e.g.,][]{Jelic2010,Bernardi2010,Moore2013,Martinot2018} but the case of global signal experiments has received very little attention so far \citep{Switzer2014}, in particular after the reported detection of the $21$~cm signal from the Cosmic Dawn.

The paper is organized as follow:
in section~\ref{sec:sim} we describe the contamination from polarized foregrounds in observations carried out with single dipole antennas and outline the details of our simulations, in section~\ref{sec:bayes} we describe the extraction of the 21~cm global signal from the simulated spectra and we conclude in section~\ref{sec:con}.

\section{Simulations of global signal observations}\label{sec:sim}
An individual antenna provides a measurement of the beam-averaged sky brightness temperature $T(\hat{\boldsymbol{r}}_0,\nu,t)$ at the time $t$ and direction $\hat{\boldsymbol{r}}_0$ \citep[e.g.,][]{Bernardi2015}:
\begin{equation}\label{eq:meas}
T(\hat{\boldsymbol{r}}_0,\nu,t) = \frac{\int_{\Omega} A(\hat{\boldsymbol{r}}',\nu) \, T_{\rm sky}( \hat{\boldsymbol{r}}',\nu,t) \, d\hat{\boldsymbol{r}}'}{\int_{\Omega} A(\hat{\boldsymbol{r}}',\nu) \, d\hat{\boldsymbol{r}}'} + T_{N}(\nu,t)
\end{equation}
where $T_{\rm sky}$ is the sky brightness temperature,
$A$ the antenna gain pattern and $T_{\rm N}$ the instrumental noise. As the sky drifts over the dipole, the sky brightness changes with time whereas the dipole pattern does not. 

A single-polarization antenna inevitably measures polarized emission from the sky.
If we call ${\bf s}$ the intrinsic sky brightness distribution towards a line of sight $\hat{\boldsymbol{r}}$ at the frequency $\nu$ in terms of the usual Stokes parameters ${\bf s}=(I,Q,U,V)^T$, the brightness observed by two orthogonal receptors ${\bf e}=(E_{xx},E_{xy},E_{yx},E_{yy})^T$
can be written as \citep[e.g.,][]{Ord2010,Nunhokee2017}:
\begin{equation}\label{eq:intfresp}
{\bf e} (\hat{\boldsymbol{r}}, \nu)		=  \big[ {\bf J}(\hat{\boldsymbol{r}}, \nu) \otimes {\bf J}^* (\hat{\boldsymbol{r}}, \nu) \big] \, {\bf S} \, {\bf s} (\hat{\boldsymbol{r}}, \nu),
\end{equation}
where ${\bf{J}}$ is the $2 \times 2$ Jones matrix representing the polarized receptor response (i.e., the polarized dipole gain pattern), $\otimes$ is the outer product operator, $^*$ denotes the complex conjugate and ${\bf{S}}$ is the matrix that relates the Stokes parameters to the orthogonal $x-y$ linear feed frame:
\begin{equation*}
{\bf{S}}=\frac{1}{2}\begin{pmatrix} 1 & 1 & 0 & 0 \\
0 &  0 & 1 &  i \\
0 &  0 & 1 & -i \\
1 & -1 & 0 &  0 \end{pmatrix}. 
\end{equation*}
The matrix ${\bf A}(\hat{\boldsymbol{r}}, \nu) \equiv \big[ {\bf J}(\hat{\boldsymbol{r}}, \nu) \otimes {\bf J}^* (\hat{\boldsymbol{r}}, \nu) \big] {\bf S}$ can be seen as a mixing matrix between the intrinsic and the observed Stokes parameters \citep[e.g.,][]{Nunhokee2017}.
A single polarization antenna is described by a Jones matrix of the form:
\begin{equation*}
{\bf{J}}=\begin{pmatrix} 
J_x & 0 \\
0 & 0 \\\end{pmatrix}, 
\end{equation*}
and equation~\ref{eq:intfresp} leads to:
\begin{equation}\label{eq:IQ}
E_{xx} (\hat{\boldsymbol{r}}, \nu)	 = \frac{1}{2} J_x^2 (\hat{\boldsymbol{r}}, \nu) \big[I(\hat{\boldsymbol{r}}, \nu,t) + Q(\hat{\boldsymbol{r}}, \nu,t)\big]. 
\end{equation}
Similarly, the orthogonal polarization would be:
\begin{equation}\label{eq:ImQ}
E_{yy} (\hat{\boldsymbol{r}}, \nu)	 = \frac{1}{2} J_y^2 (\hat{\boldsymbol{r}}, \nu) \big[I(\hat{\boldsymbol{r}}, \nu,t) - Q(\hat{\boldsymbol{r}}, \nu,t)\big].
\end{equation}
By renaming $A_{x,y}\equiv \frac{1}{2}J^2_{x,y}$, 
equation~\ref{eq:meas} can be re-written explicitly for both polarizations:
\begin{eqnarray}\label{eq:pQmQ}
T_{xx}( \hat{\boldsymbol{r}}_0,\nu,t) = \frac{\int_{\Omega} E_{xx} (\hat{\boldsymbol{r}}', \nu) d\hat{\boldsymbol{r}}'}{\int_{\Omega} A_x(\hat{\boldsymbol{r}}',\nu)d\hat{\boldsymbol{r}}'} = T_{f}( \hat{\boldsymbol{r}}_0,\nu,t) + T_Q( \hat{\boldsymbol{r}}_0,\nu,t) + T_{21}(\nu) \nonumber \\
T_{yy}( \hat{\boldsymbol{r}}_0,\nu,t) = \frac{\int_{\Omega} E_{yy} (\hat{\boldsymbol{r}}', \nu) d\hat{\boldsymbol{r}}'}{\int_{\Omega}  A_y(\hat{\boldsymbol{r}}',\nu)d\hat{\boldsymbol{r}}'} = T_{f}( \hat{\boldsymbol{r}}_0,\nu,t) - T_Q( \hat{\boldsymbol{r}}_0,\nu,t) + T_{21}(\nu)
\end{eqnarray}
where $T_f$ and $T_Q$ are the foreground contribution from intensity and polarization, respectively (examined in section~\ref{sec:fg} and \ref{sec:pol}) and $T_{21}$ is the contribution to the sky brightness coming from the pristine $21$~cm signal that we will discuss further in section~\ref{sec:signal}.
Note that we have here neglected the contribution from $21$~cm fluctuations as it essentially averages out over large sky areas.

Our goal is to simulate an observed spectrum ${\bar T}_{xx,yy}$ obtained by averaging $T_{xx,yy}$ over the observing time - i.e. the data product of a global signal experiment:
\begin{eqnarray}\label{eq:Tmean}
\bar{T}_{xx}(\nu) = \frac{1}{N_m}\sum{T_{xx}(t,\nu)} = \bar{T}_f(\nu) + \bar{T}_Q(\nu) + T_{21}(\nu) \nonumber \\ 
\bar{T}_{yy}(\nu) = \frac{1}{N_m}\sum{T_{yy}(t,\nu)} = \bar{T}_f(\nu) - \bar{T}_Q(\nu) + T_{21}(\nu),
\end{eqnarray}
where $N_m$ is the number of measurements over the observation duration. 

We consider a dipole located at the Murchison Radio-astronomy Observatory in Western Australia,  where EDGES is located, and that observes the $0^{\rm h} < {\rm LST} < 8^{\rm h}$~hour range with a one minute cadence.
We assume that the noise $T_N$ is given by the radiometer equation: it is uncorrelated in frequency and time, and, for each frequency channel, follows a Gaussian distribution with standard deviation 
\begin{equation}\label{eq:noise}
\sigma^N_{xx,yy}(\nu)=\frac{\bar{T}_{xx,yy}(\nu)}{\sqrt{\Delta t \Delta \nu}},
\end{equation}
where we consider a $\Delta \nu = 1$~MHz channel width and a $\Delta t = 400$~hours of total integration time. 
Like EDGES, we consider two separate bands, one covering the low frequency (LF) $50-100$~MHz range, and the second covering the higher frequency (HF) $100-200$~MHz range.

\subsection{Antenna beam model}\label{sec:beam}
\begin{figure}
\includegraphics[width=\columnwidth]{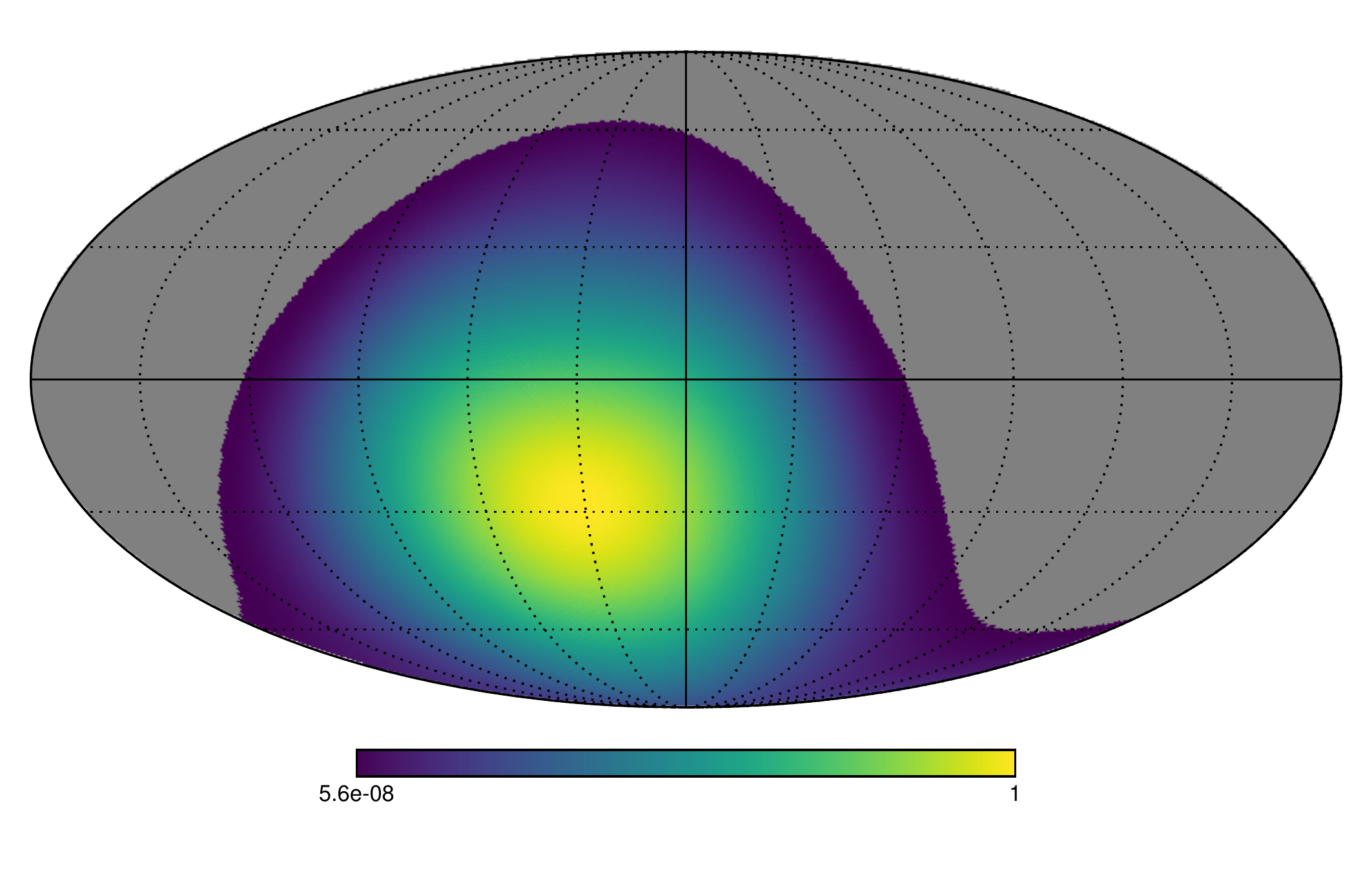}
\includegraphics[width=\columnwidth]{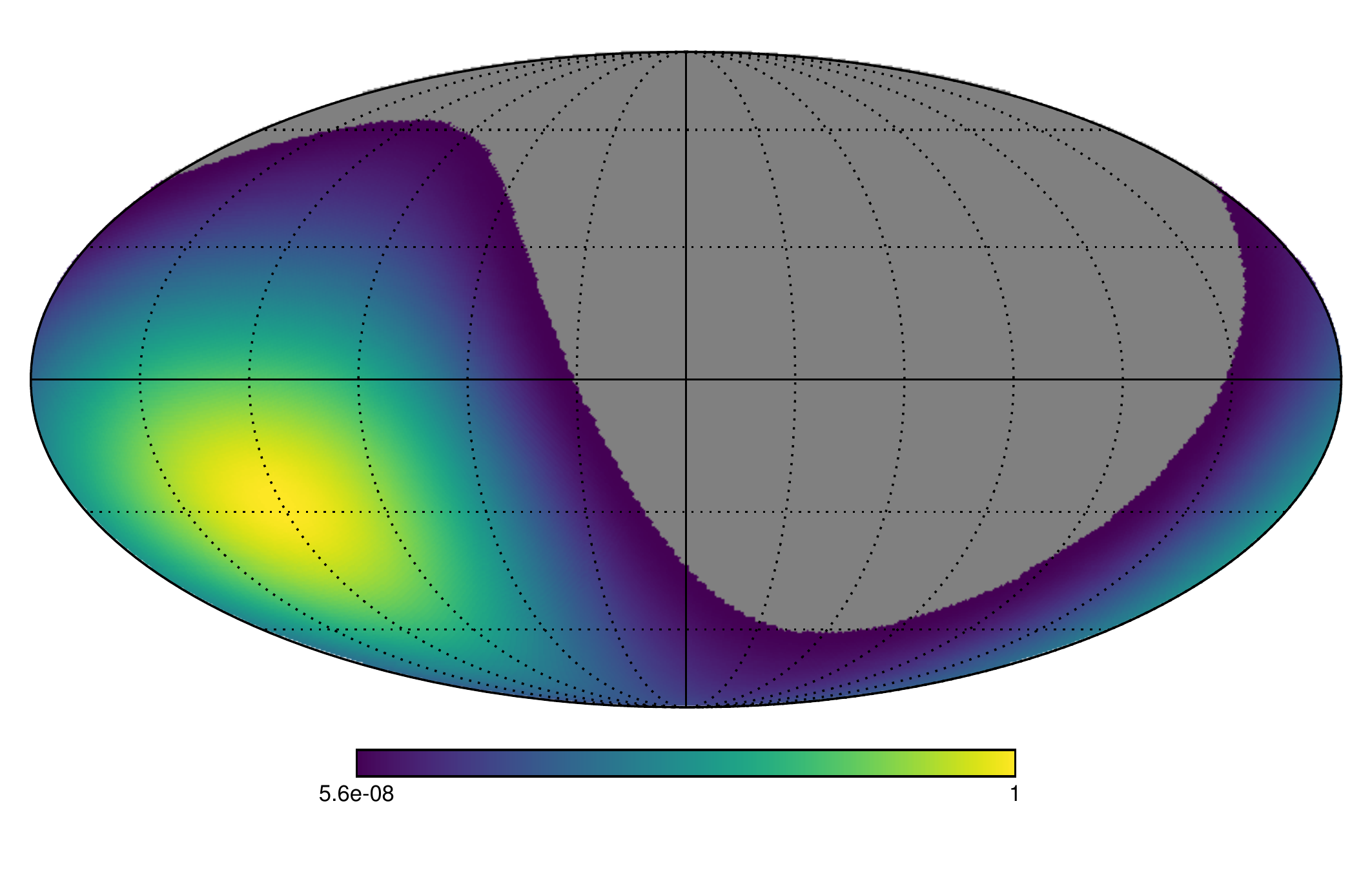}
\includegraphics[width=\columnwidth]{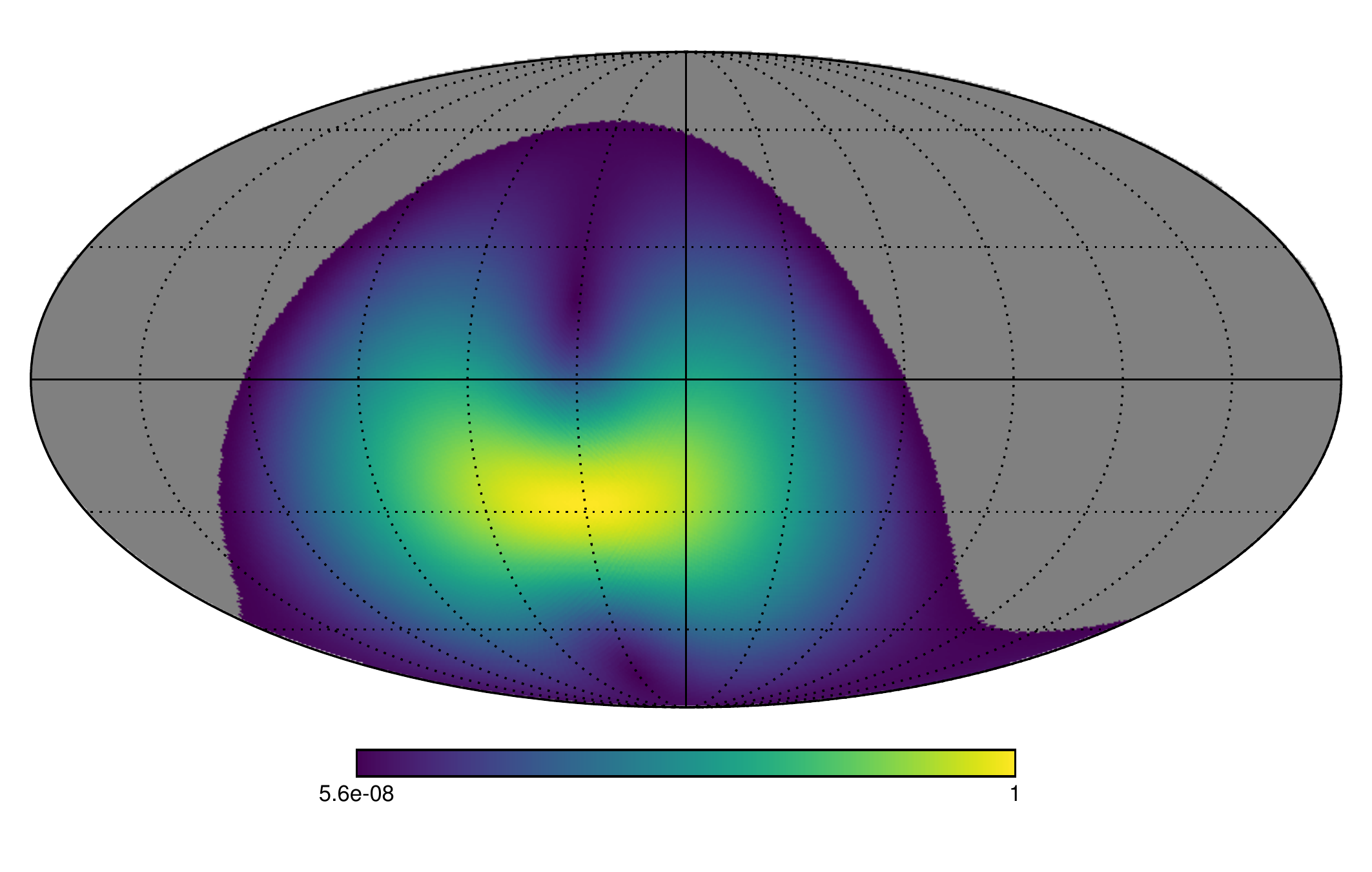}
\caption{E-W ($xx$) dipole beam model at $50$~MHz for $\mathrm{LST}=2^{\rm h}$ (top panel) and $\mathrm{LST}=8^{\rm h}$ (middle panel). The bottom panel shows instead the $100$~MHz beam again for $\mathrm{LST}=2^{\rm h}$.}\label{fig:beam}
\end{figure}

We used the analytic beam model of the Long Wavelength Array dipole \citep{Taylor2012,Ellingson2013,Bernardi2015} in the LF band:
\begin{equation}
A(\theta,\phi,\nu)=\sqrt{[p_E(\theta,\nu) \cos{\phi}]^2 + [p_H(\theta,\nu) \sin{\phi}]^2},
\notag
\end{equation}
where $E$ and $H$ are the two orthogonal polarizations of the dipole
and
\begin{equation}
p_i(\nu,\theta)=\left[ 1 - \left( \frac{\theta}{\pi/2} \right)^{\alpha_i(\nu)} \right](\cos{\theta})^{\beta_i(\nu)} + \gamma_i(\nu) \left(\frac{\theta}{\pi/2} \right)(\cos\theta)^{\delta_i(\nu)}
\end{equation}
where $i=E, H$. For the coefficient $[\alpha_i,\beta_i,\gamma_i, \delta_i]$ we use the values tabulated in \citet{Dowell2011} and interpolate them in the $50-90$~MHz range. The values of the coefficients are then extrapolated to $100$~MHz with a $3^{\rm rd}$-order polynomial. Figure~\ref{fig:beam} displays the beam model for the E-W ($xx$) orientation at $\mathrm{LST}=2^{\rm h}$, at 50 and~100 MHz respectively.
For modelling the N-S (yy) orientation we switch the E and H terms.

In the absence of a publicly available beam model in the HF band, we directly scale our 100~MHz model linearly with frequency up to 200~MHz.

\subsection{Global signal model}\label{sec:signal}
\begin{figure}
\includegraphics[width=\columnwidth]{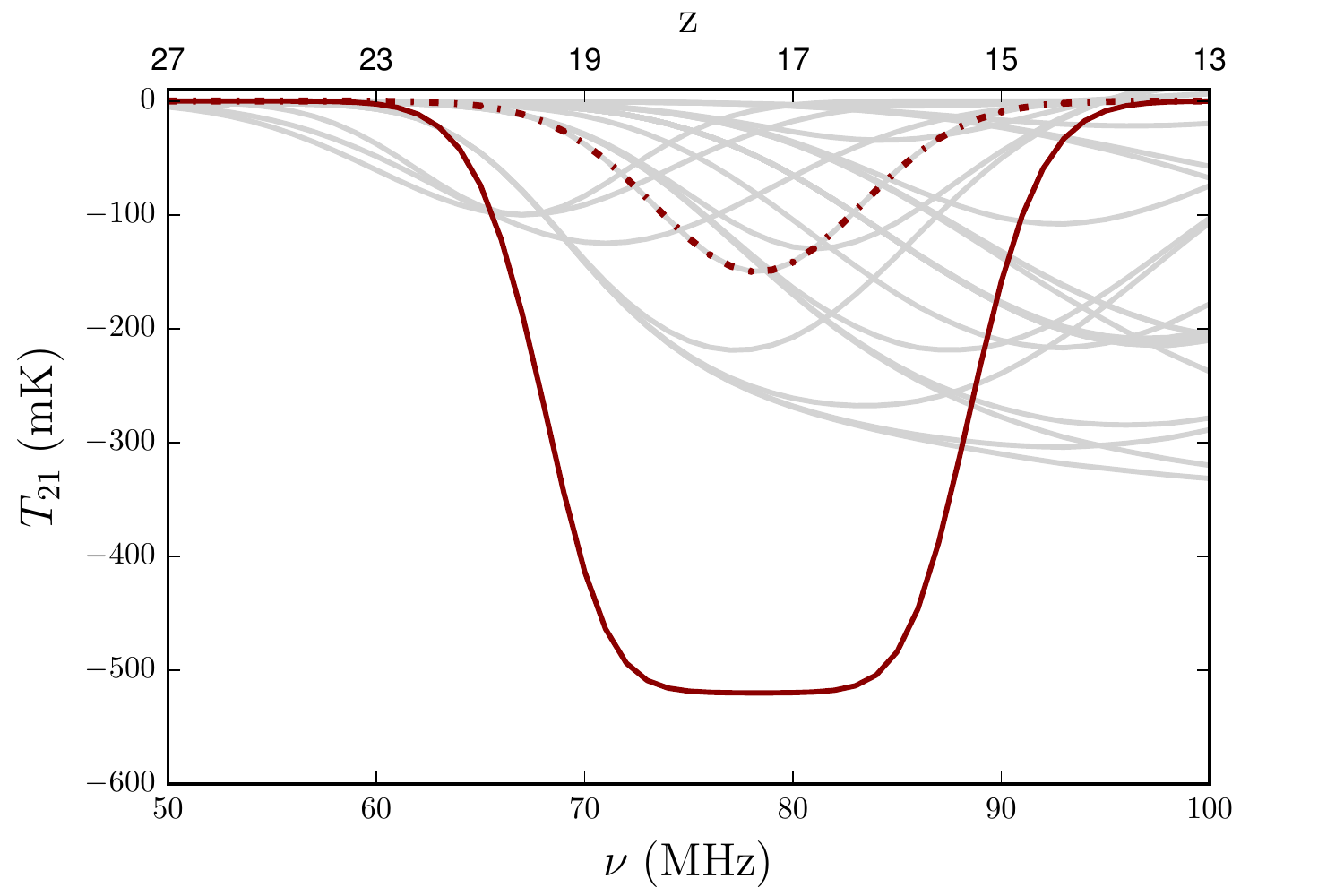}
\caption{Our fiducial Gaussian input model (dot-dashed red line) and the flattened Gaussian model best-fit of the EDGES data \citep[][solid red line]{EDGES},  compared with the global signal profiles obtained with {\it SimFast21} \citep{Santos2010} varying the physical input parameters (solid grey lines).}
\label{fig:input_theory}
\end{figure}

The evolution of $21$~cm global signal can be computed from physical model parameters via numerical or semi-analytical simulations  \citep[e.g.,][]{Mirocha2014,Mirocha2015,Cohen2016,Cohen2017,Mirocha2017}, however, analytic expressions are useful approximation to be used in the evaluation of likelihood functions.
In the LF band, the Cosmic Dawn signal has often been modelled as a Gaussian absorption profile \citep{Bernardi2015,Presley2015,Bernardi2016,Monsalve2017}:
\begin{equation}\label{eq:gauss}
T_{21,G}^{LF}(\nu)=A_{21}e^{-\frac{(\nu-\nu_{21})^2}{2\sigma_{21}^2}},
\end{equation}
where $A_{21}$, $\nu_{21}$ and $\sigma_{21}$ are the amplitude, peak position and standard deviation of the $21$~cm trough, respectively. We consider this our fiducial model for the LF band.
We also include the case of a flattened Gaussian profile adopted in the EDGES analysis \citep{EDGES}:
\begin{equation}\label{eq:flat}
T^{LF}_{21,fG}(\nu)=A_{21} \left( \frac{1-e^{-\tau e^B}}{1-e^{-\tau}}\right),
\end{equation}
where 
\begin{equation}
B=\frac{4(\nu -\nu_{21})^2}{w^2}\log{\left[ -\frac{1}{\tau} \log{\left(\frac{1+e^{-\tau}}{2}\right)} \right]}.
\end{equation}
The free parameters here are the amplitude $A_{21}$, the central frequency $\nu_{21}$, the full-width at half-maximum $w$ and the flattening factor $\tau$.
Theoretical simulations that include standard physics predict a wide range of different global 21~cm signals (see, fore example, figure~\ref{fig:input_theory}). The largest theoretical unknown is related to the nature of the first luminous sources \citep[e.g.,][]{Furlanetto2006,Mirocha2014} and the efficiency of the IGM heating \citep{Pritchard2007,Fialkov2013,Mesinger2013,Cohen2017}. As shown in figure~\ref{fig:input_theory}, not even models that predict the brightest absorption profiles are a close match to the EDGES result.
Our fiducial model in the HF band is an hyperbolic tangent, a widely used parameterization of the global signal during EoR
\citep[e.g.,][]{Pritchard2010,Monsalve2017}:
\begin{equation}\label{eq:hf}
T^{HF}_{21}(z)=a_{21}x_{HI}(z)\sqrt{\frac{1+z}{10}} 
\end{equation}
where $a_{21}=28$~mK \citep{Madau1997,Furlanetto2006} and
\begin{equation}
x_{\rm HI}(z)=\frac{1}{2}\left[ \tanh{\left(\frac{z-z_r}{\Delta z}\right)} + 1 \right].
\end{equation}
The free parameters are here the redshift $z_r$ at which $x_{\rm HI} = 0.5$ and the reionization duration, $\Delta z = (dx_{HI}/dz)^{-1}|_{x_{\rm HI}=0.5}$.

\subsection{Total intensity foreground model}\label{sec:fg}

A total intensity all-sky map could be used to evaluate the observed foreground spectrum ${\bar T}_f$ via equation~\ref{eq:pQmQ} and \ref{eq:Tmean} as it was done, for example, in \citet{Bernardi2015}. Rather than repeating a similar simulation, we directly calculated the total intensity foreground spectrum averaged over the duration of the observations, i.e. the left hand side of equation~\ref{eq:Tmean}.

The Galactic foreground spectrum has often been modelled as a $N^{\rm th}$-order log-polynomial \citep[e.g.,][]{Bowman2010,Pritchard2010,Harker2012,Bernardi2015,Presley2015,Bernardi2016}:
\begin{equation}\label{eq:fg}
\log_{10}{\bar{T}_f(\nu)} = \sum_{n=1}^{N} p_{n-1} \left[ \log_{10}{ \frac{\nu}{\nu_0}} \right]^{(n-1)}
\end{equation}
with $\nu_0=60$~MHz.
In earlier works, the foreground spectrum was modelled with few frequency components \citep[e.g.,][]{Pritchard2010}, but more recent simulations suggest that, due to the coupling between the antenna beam pattern and the sky brightness, $N$ should likely take higher values \citep{Harker2012,Bernardi2015,Bernardi2016,Mozdzen2016}.
Here we used the best fit coefficients derived from simulations in \citep{Bernardi2015}, with a $N=4$ log-polynomial (see Table~\ref{tab:fg}), a case similar to the analysis in \citet{EDGES}. 

\begin{table}\caption{Coefficients of the Galactic synchrotron spectrum model \citep[from][]{Bernardi2015}.}\label{tab:fg}\begin{tabular}{cccccccc}
$\log_{10}{(p_0/{\rm K})}$ & $p_1$ & $p_2$ & $p_3$ & $p_4$ \\ \hline
3.58 & -2.60 & 0.01 & 0.06 & 0.25 \\\hline
\end{tabular}\end{table}

\subsection{Polarized foreground model}\label{sec:pol}

\begin{figure}
\includegraphics[width=\columnwidth]{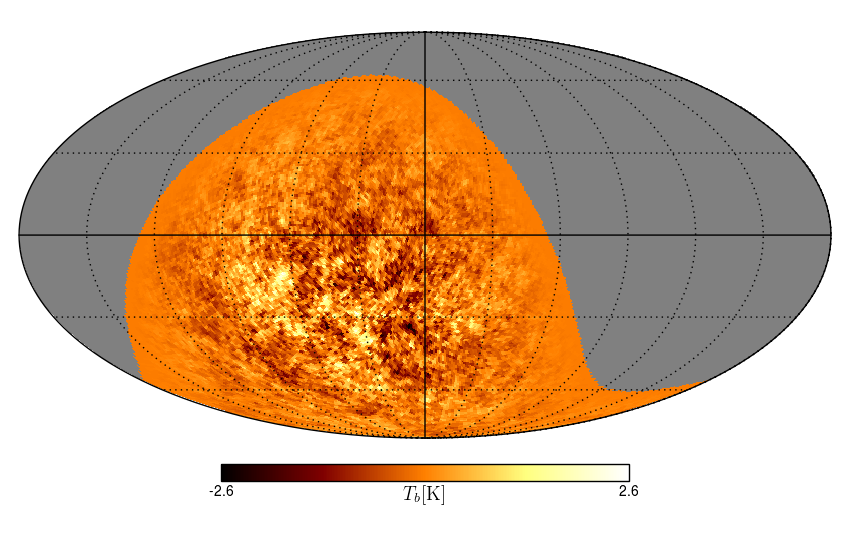}
\caption{Example of a simulated Stokes~$Q$ map at 80~MHz after the dipole beam pattern is applied.}\label{fig:sky}
\end{figure}

\begin{figure}
\includegraphics[width=\columnwidth]{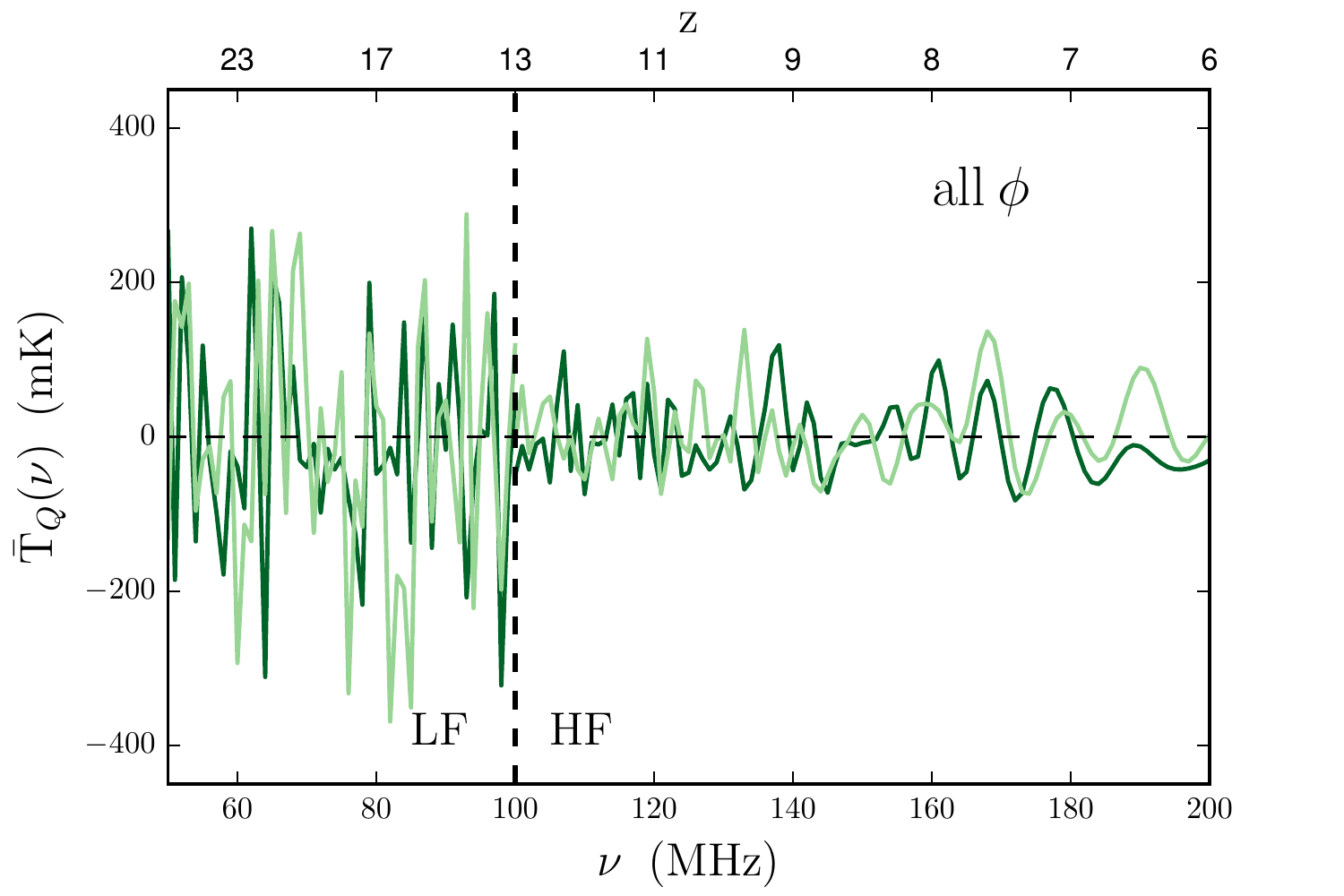}
\includegraphics[width=\columnwidth]{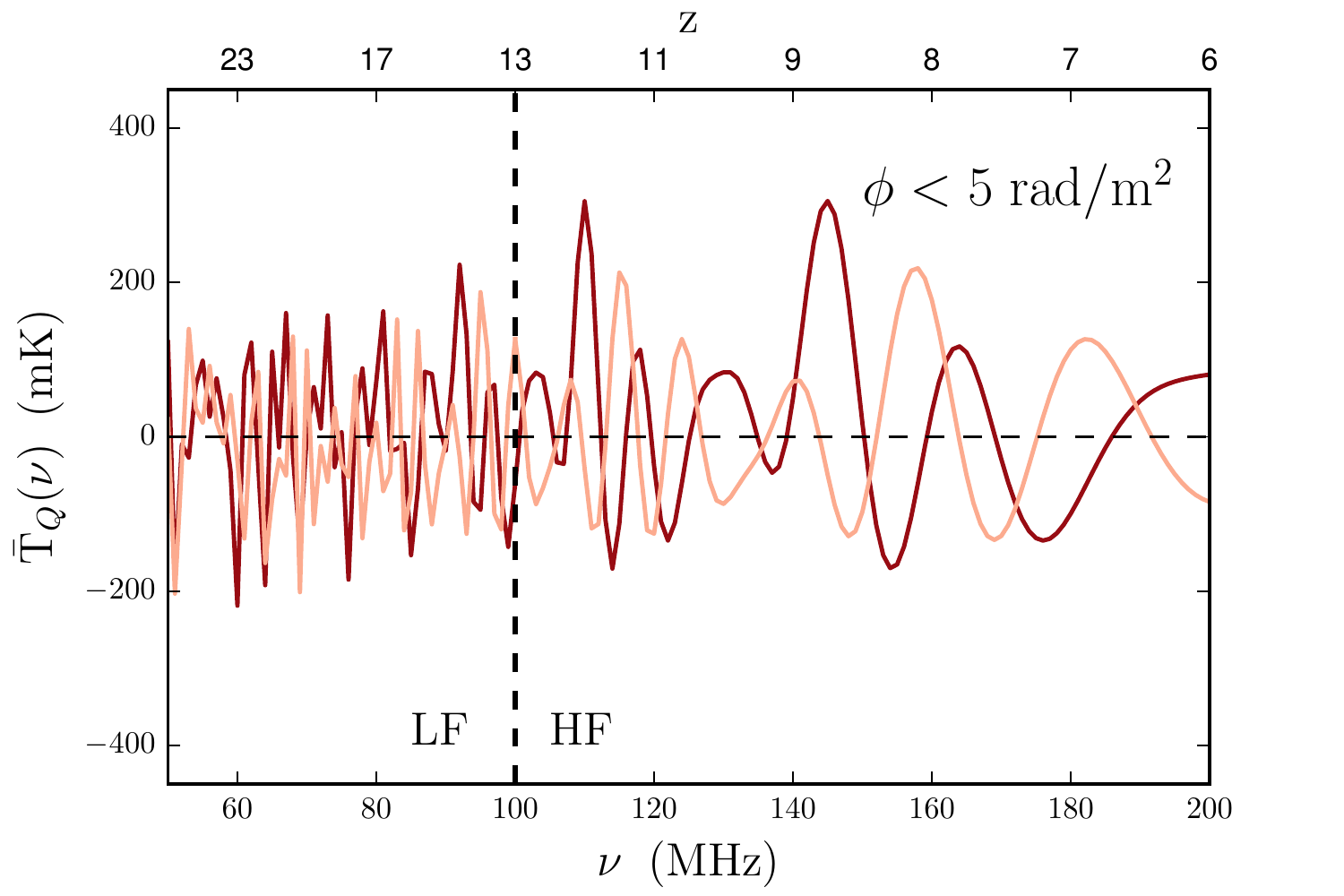}
\caption{{\it Top panel}: simulated polarized foreground spectra $\bar{T}_Q(\nu)$ (equation~\ref{eq:meas}) integrated over the LST range. Light and dark green solid lines correspond to two different realizations of the ``all $\phi$" simulations. The vertical dashed line divides the LF band from the HF one. {\it Bottom panel}: same as the top panel but with light and dark red lines corresponding to two different realizations of the ``low $\phi$" (i.e. $\phi < 5$~rad~m$^{-2}$) simulations.}\label{fig:Tvsnu}
\end{figure}

We use the simulations in \citet[][hereafter S18]{PolSynch} to produce Stokes~$Q$ and $U$ full sky maps in the 50-200~MHz range with 1~MHz frequency resolution. 
The S18 full sky simulations are based on the interferometric observations that sample up to degree angular scales \citep{Bernardi2013} that were extrapolated up to tens of degrees scales, relevant for global signal observations.
They are constructed from rotation measure  synthesis data that measure the polarized intensity as a function of Faraday depth $\phi$ \citep{Burn1966,Brentjens2005}.
Figure~\ref{fig:sky} displays an example of a Stokes $Q$ map observed through the dipole beam (equation~\ref{eq:pQmQ}).
We generate two sets of polarized foreground spectra $\bar{T}_{Q} (\nu)$:
\begin{enumerate}
\item one that uses S18 simulations with the full range of $\phi$ values from the data. We will refer to this simulation as the ``all $\phi$" case;
\item a second one where high values of the Faraday depth $\phi$ ($\phi > 5$~rad/m$^{2}$) are excluded from the S18 simulations. The motivation behind this choice is to create a more realistic model in the LF band. Observations indicate that Galactic polarized emission has a more local origin with decreasing frequency \citep[e.g.,][]{Haverkorn2004,Bernardi2009,Lenc2016} and, therefore, very little emission at high Faraday depth values. We will refer to this simulation as the ``low $\phi$" case (i.e. $\phi < 5$~rad/m$^{2}$). 
\end{enumerate}
The combination of the integrated effect of the beam and the complex Faraday structure result in spectra like the one shown in figure~\ref{fig:Tvsnu}, where 
two representative realizations of both sets of simulations
for both the LF and HF band are displayed.  
In the LF band, the ``all $\phi$" simulation leads to a considerable more complex spectral structure and higher contamination with respect to the ``low $\phi$" case - as expected.
The oscillatory behaviour becomes smoother in the HF band for both cases but, although there are fewer peaks, the contamination is more prominent for the ``low $\phi$" case.

\begin{figure}
\includegraphics[width=\columnwidth]{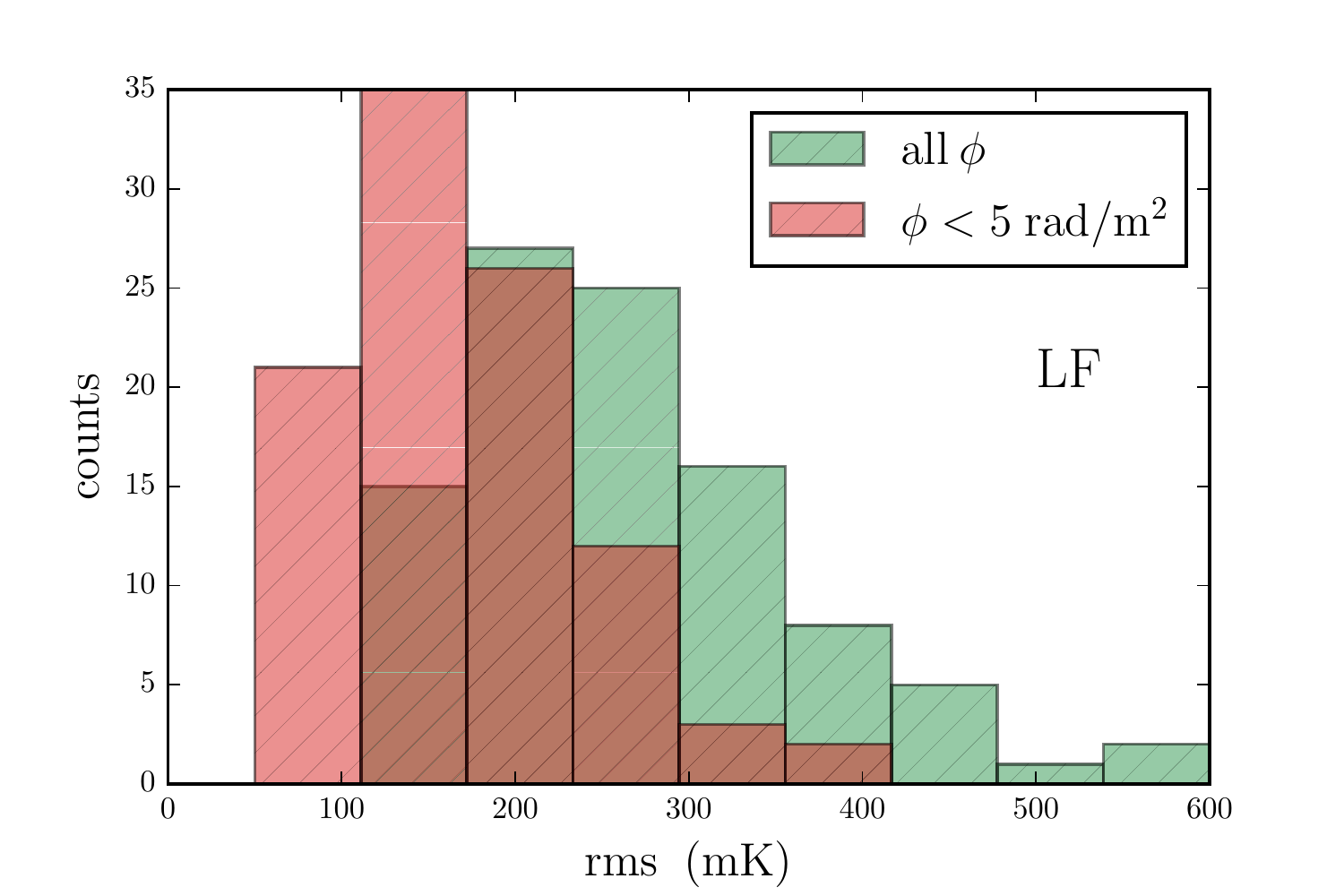}
\includegraphics[width=\columnwidth]{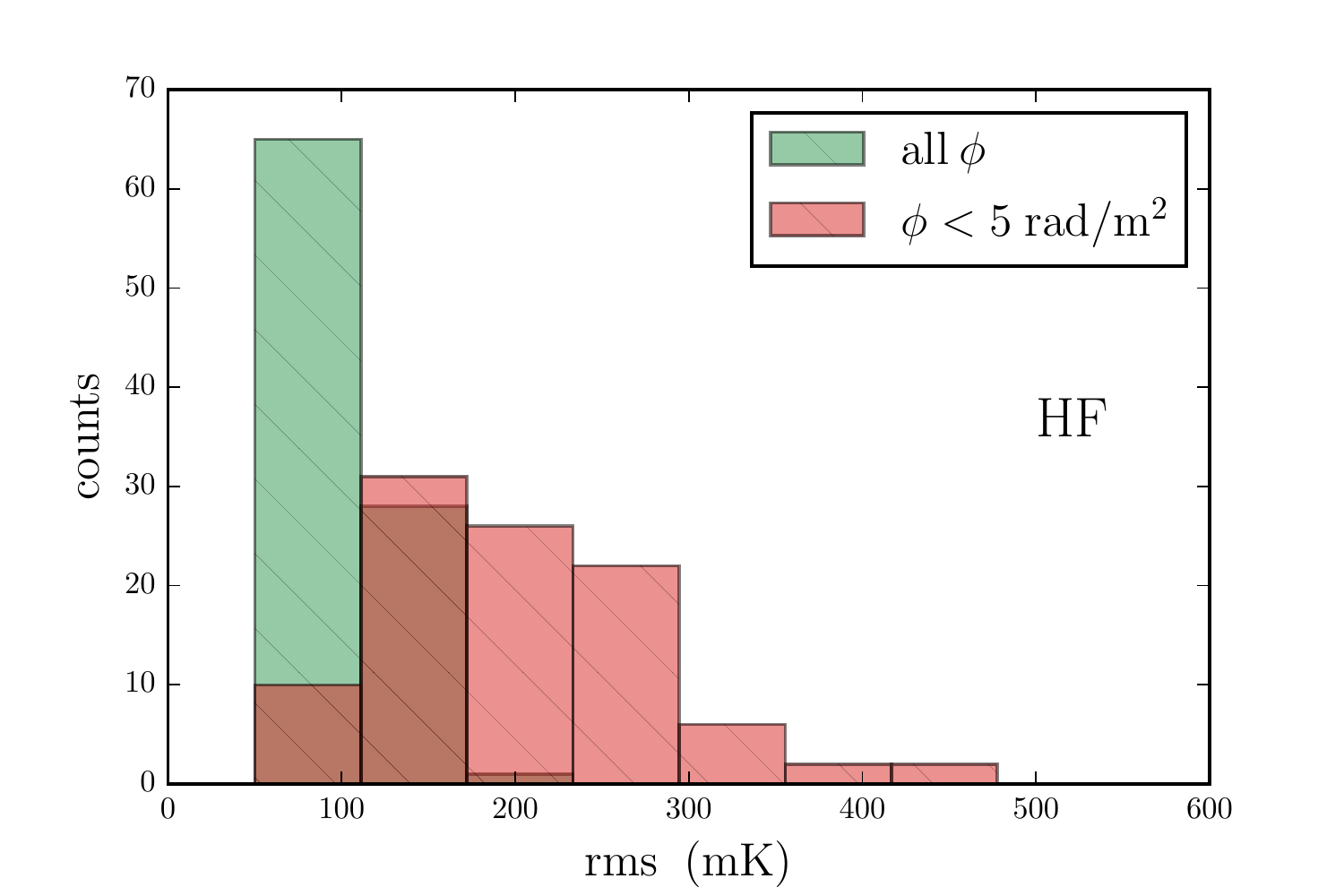}
\caption{{\it Top panel}: distribution of the polarized spectrum rms calculated in the LF band from 100 realizations for both the ``all $\phi$" (in green) and the ``low $\phi$" (i.e. $\phi < 5$~rad/m$^{2}$) simulations (in red). {\it Bottom panel}: same as the top panel but for the HF band.}
\label{fig:hist}
\end{figure}

We calculated the rms of $\bar{T}_{Q} (\nu)$ for every realization and plot its distribution in figure~\ref{fig:hist}.
In the LF band, the average rms contamination is $\sim 250$~mK for the ``all $\phi$" case, with an extended tail at high values. The average contamination is smaller in the other case, peaking around $\sim 150$~mK.
The situation is opposite in the HF band, where the ``all $\phi$" simulation has an average rms contamination smaller than $\sim 100$~mK whereas the ``low $\phi$" case spans a much broader range of values with an extended tail up to $\sim 400$~mK. It is worth noticing that the estimated rms contamination is generally higher or comparable with the expected $21$~cm signal, although our simulations likely represent a worst case scenario as they do not account for time of frequency dependent depolarization effects. Time variable electron density and variations of the magnetic fields across the field of view both depolarize the signal when integrated over long observations. Simulations by \cite{Martinot2018} estimated the depolarization to be a factor of four or more when averaging over days and over a $\sim 10^\circ$ sky patch. The effect can be even more pronounced for global signal observations. Frequency dependent polarization arises when emitting clouds are Faraday thick, i.e. synchrotron emission and Faraday rotation are co-located within the cloud \citep{Burn1966,Tribble1992}. Its magnitude depends upon the detailed physics of the interstellar medium and therefore it is fairly uncertain. We note, however, that polarized fluctuations at 350~MHz are of the order of a few Kelvin that, extrapolated at 150~MHz with a fiducial spectral index $\beta = -2.6$ would lead to polarized signals at the level of a few tens of Kelvin. Polarized fluctuations remain at the $10-20$~K level in the $150-200$~MHz range \citep[e.g.,][]{Bernardi2013,Lenc2016}, implying that part of the emission happens in Faraday thick regions and it is Faraday depolarized at low frequencies. In order to empirically account for these effects, we also considered a more optimistic case where the magnitude of the polarized spectrum is reduced to a $10\%$ of the current simulation value. This choice is in qualitative agreement with the magnitude of the residual rms in the \citet{EDGES} observations.

The final product of our simulations is a sky spectrum $\bar{T}(\nu)$ that is the sum of four different components: a $21$~cm signal $T_{21}(\nu)$ as described in section~\ref{sec:signal}; a total intensity foreground spectrum that follows a $N^{th}$-order log polynomial (see section~\ref{sec:fg}); a polarized foreground spectrum $\bar{T}_Q(\nu)$ (section~\ref{sec:pol}) and a noise realization drawn from a Gaussian distribution  (equation~\ref{eq:noise}). 
The next section describes the extraction of the $21$~cm signal from the simulated spectra. 


\begin{figure*}
\includegraphics[width=15cm]{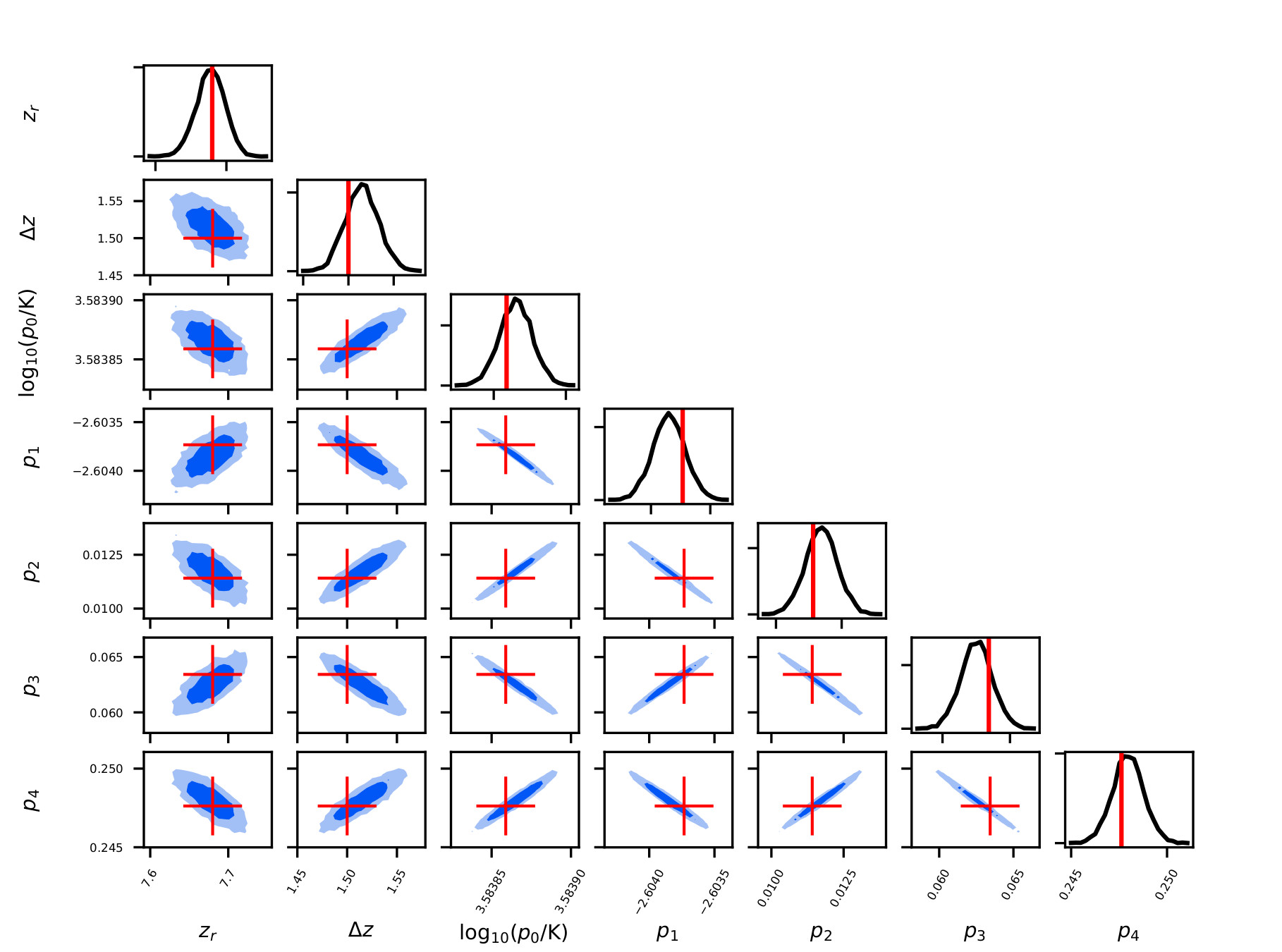}
\caption{Marginalized two dimensional posterior distributions of the global $21$~cm signal in the HF band and only the total intensity foreground parameters, without including any contamination from polarized foregrounds. Contours are shown at $1\sigma$ and $2\sigma$ respectively, whereas the red crosses indicate the input parameter values. The agreement shows that the reconstruction is unbiased in the case of smooth foregrounds.}\label{fig:bayes}
\end{figure*}

\section{Signal extraction}\label{sec:bayes}

In order to extract the global $21$~cm signal from the simulated spectra we use the \textsc{hibayes} code \citep{Bernardi2016,Zwart2016}, a fully Bayesian framework where the posterior probability distribution is explored through the \textsc{multinest} sampler \citep{Feroz2008,Feroz2009} using an MPI-enabled python wrapper \citep{Buchner2014}. 
The likelihood $\mathcal{L}_i$ of the simulated spectra can be written as:
\begin{equation}
 \mathcal{L}(\bar{T}(\nu_i)|\boldsymbol{\theta}) =\frac{1}{\sqrt{2\pi \sigma_N^2(\nu_i)}}\exp{\left(-\frac{(\bar{T}(\nu_i)-T_m(\nu_i,\boldsymbol{\theta}))^2}{2\sigma_N^2(\nu_i)}\right)},
\end{equation}
where $\boldsymbol{\theta}$ is the vector of model parameters, $\sigma_N$ is the noise standard deviation (equation~\ref{eq:noise}) and $T_m(\nu_i,\boldsymbol{\theta})$ is the model spectrum.
We impose uniform prior on the signal parameters assuming the signal is present within the observed band. For the HF band this translates into a limit for the middle point of reionization i.e. $6<z_r<13$ and for the reionization duration i.e. $0<\Delta z < 13$. In the LF band we set the priors to be $40<\nu_{21}<100$~MHz, $0<\sigma_{21}<50$~MHz and, solely to reduce the computational load, $-1<A_{21}<0$~K. 
We use uniform priors for all the foreground parameters but for the $p_{n=0}$ case where we use a flat logarithmic prior. 

As a test case similar to the simulations carried out in \citet{Harker2012} and \citet{Bernardi2016}, we show in figure~\ref{fig:bayes}, the recovery of the global $21$~cm signal in the HF band (equation~\ref{eq:hf}) with $z_r=7.68$ and $\Delta z=1.50$, in agreement with \citet{PlanckXIII} and in the analysis by \citet[][]{Monsalve2017}.

We then add the simulated polarized spectrum to the total intensity one. We simulate both equation~\ref{eq:IQ} and \ref{eq:ImQ}, i.e. both the $xx$  and $yy$ polarization.
We extract the $21$~cm signal from three different simulated cases: 
\begin{itemize}
\item the 21~cm signal in the LF band is a flattened-Gaussian with $A_{21}=-520$~mK, $\nu_{21}=78.3$~MHz, $w=20.7$~MHz and the flattening parameter $\tau=7$, i.e. the EDGES best fit model \citep{EDGES}. The model spectrum used in the likelihood function is $T_m = {\bar T}_f + T_{21,fG}^{\rm LF}$. 

We generated 50 different realizations of the polarized foreground spectra and reconstruct $T_{21}(\nu)$ from the best fit parameters of the posterior distribution for each of them. We discard the cases where our reconstructed signal is localized at high frequency ($>90$~MHz) as the presence of an absorption signal in the EDGES High-Band has been excluded at $\gtrsim 2\sigma$ \citep{Monsalve2017}. After this selection, we are left with $\sim 80\%$ of the total number of simulations.
The mean and variance of the reconstructed 21~cm profiles are computed separately for the $xx$ and $yy$ polarizations and displayed as a shaded region in figure~\ref{fig:flat_flat}.

Due to the unmodelled polarized component, the residual spectra obtained after subtracting the best fit model, have relatively high rms values, at the $90 - 150$~mK level.
In the ``all $\phi$" case, the presence of an unmodelled polarized foreground introduces a bias in both the amplitude and the width of the reconstructed signal. 
In the ``low $\phi$" case the bias is mainly in the amplitude although the reconstructed flattening parameter is often different between the two polarization cases.
Figure~\ref{fig:flat_flat} shows that the reconstructed amplitude is up to $\sim 40\%$ different than the input signal, at $1\sigma$ confidence level.

\begin{figure}
\includegraphics[width=\columnwidth]{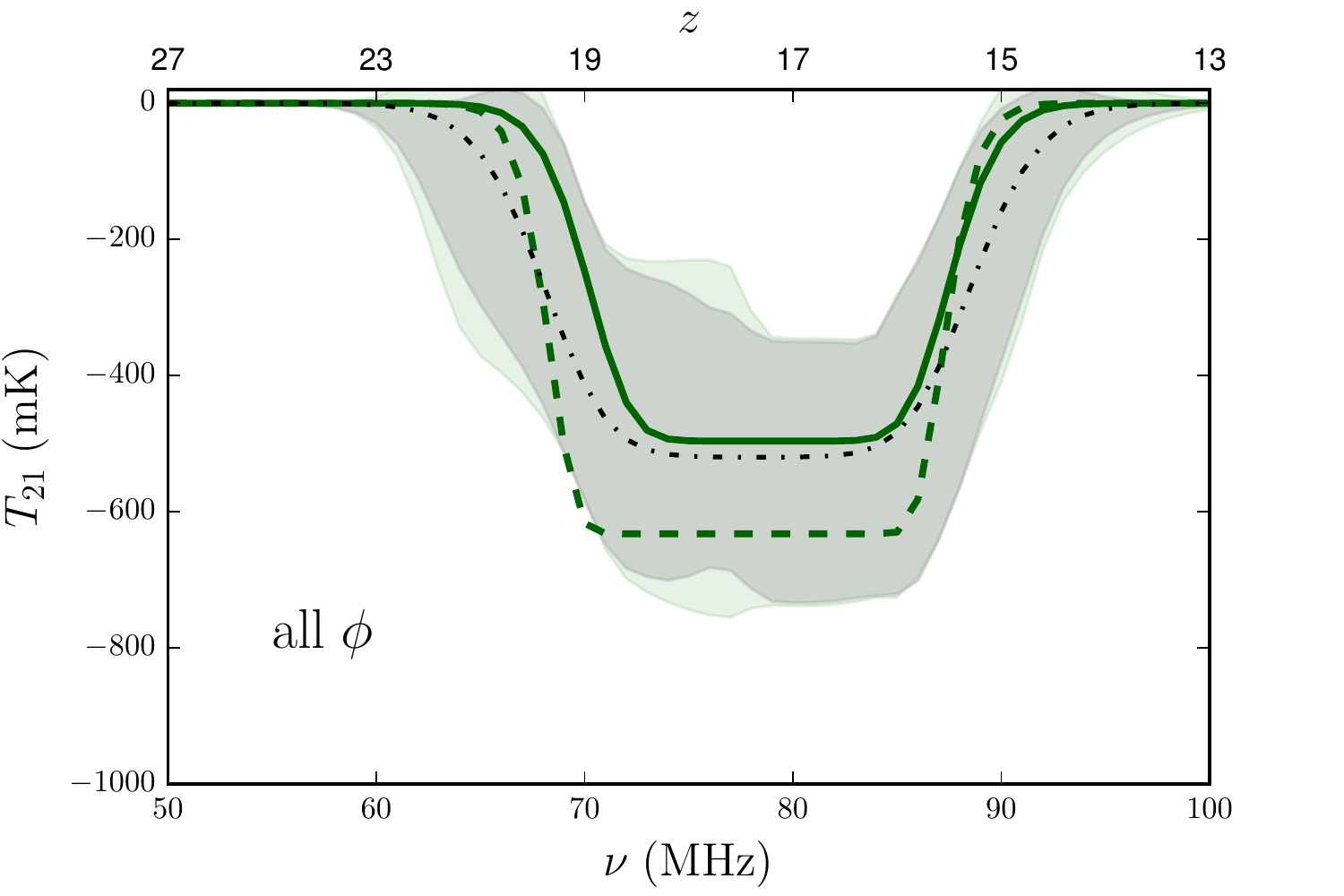}
\includegraphics[width=\columnwidth]{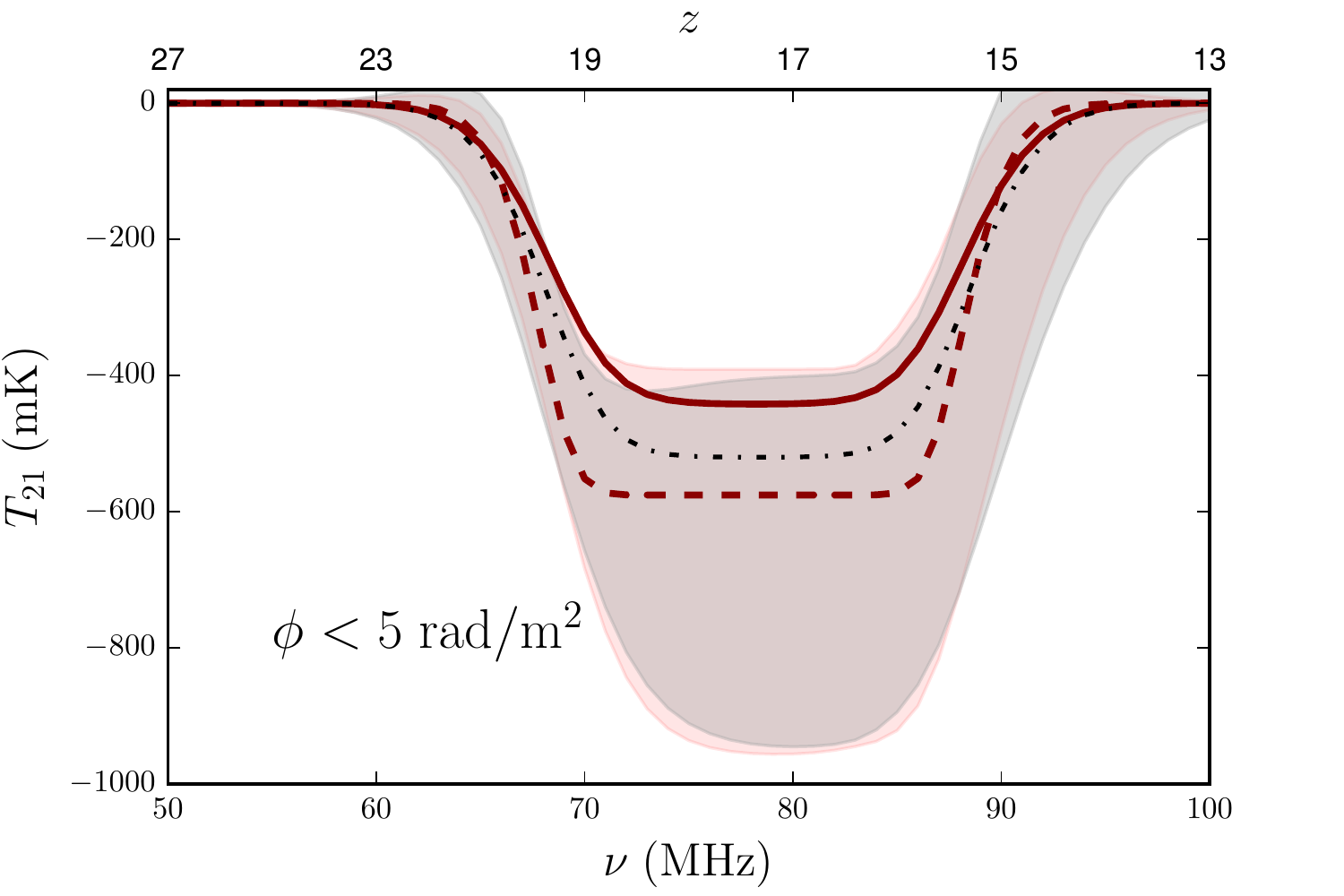}
\caption{The black dotted-dashed line in both panels is the input signal: the best fit flattened Gaussian from \citet{EDGES}. {\it Top panel}: reconstructed $T_{21,fG}$ signal in the case of the ``all $\phi$" simulations, in the LF, from the Bayesian analysis described in the text.
The solid (dashed) green line shows one of the the reconstructed $T_{21,fG}$  signals for the $xx$ ($yy$) polarization The green (grey) shaded area is the $1\sigma$ region around the mean for the $xx$ ($yy$) polarization (see text for details). {\it Bottom panel}: same but for the ``low $\phi$" (i.e. $\phi < 5$~rad~m$^{-2}$) simulations. The red (grey) shaded area is the $1\sigma$ region around the mean for the $xx$ ($yy$) polarization and the solid (dashed) red line shows one of the the reconstructed signal for both $xx$ ($yy$) case.
}\label{fig:flat_flat}
\end{figure}

\begin{figure}
\includegraphics[width=\columnwidth]{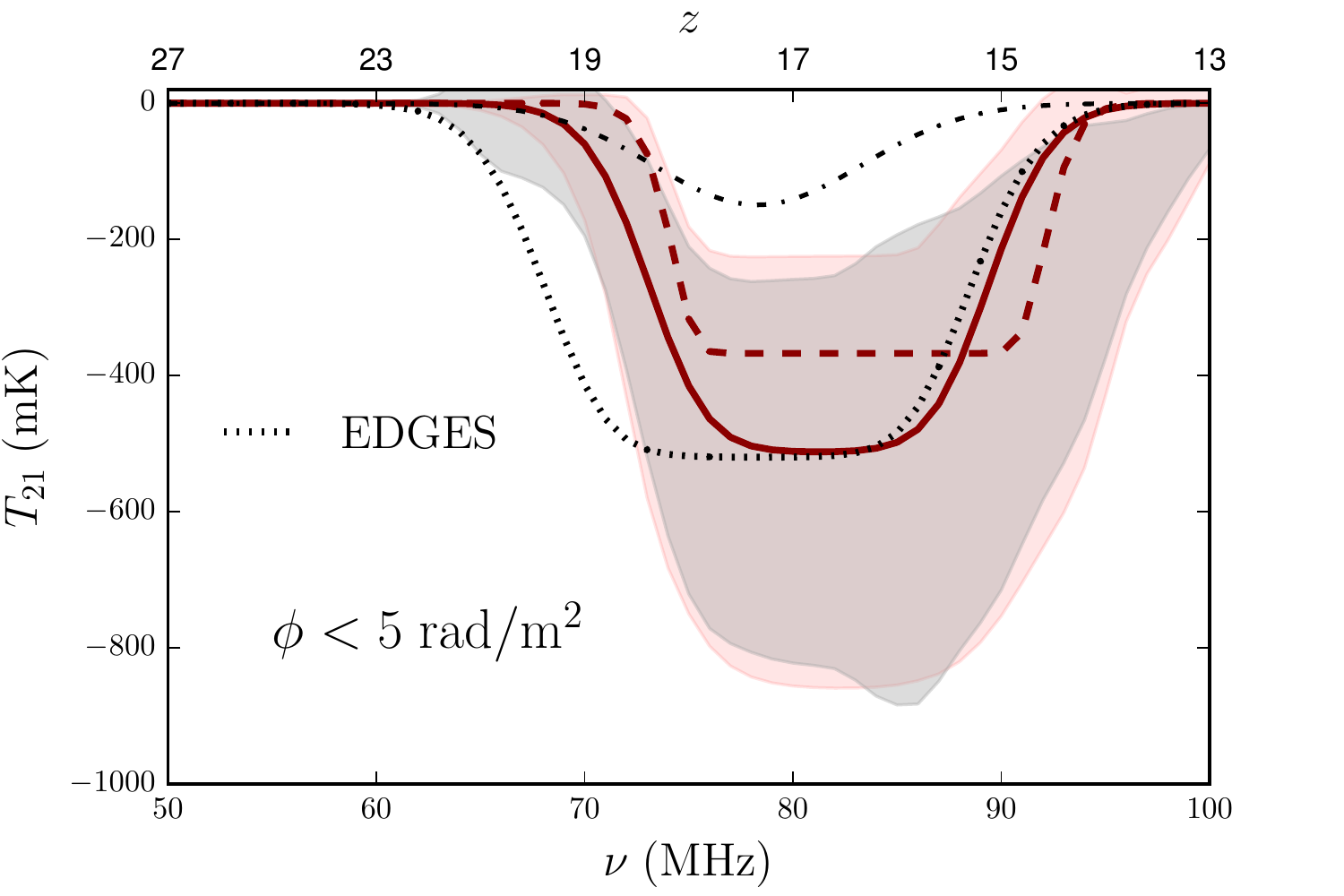}
\caption{Reconstructed $T_{21,fG}$ signal. Note that the input signal is the fiducial Gaussian model (black dotted-dashed). The solid (dashed) red line shows one of the reconstructed $T_{21,fG}$ signals for the $xx$ ($yy$) polarization. The red (grey) shaded area is the $1\sigma$ region around the mean for the $xx$ ($yy$) polarization (see text for details). For comparison we also show the EDGES best fit (dotted line). }\label{fig:gauss_flat}
\end{figure}

\item the $21$~cm signal in the LF band is a Gaussian with $A_{21}=-150$~mK, $\nu_{21}=78.3$~MHz and $\sigma_{21}=5$~MHz, i.e. the fiducial signal expected from standard theoretical models \citep[e.g.,][]{Pritchard2010,Mirocha2015}. We first model this signal using a flattened Gaussian shape in order to test whether or not the unusual shape reported by \citet{Bowman2018} can be due to the contamination from polarized foregrounds, i.e. $T_m = {\bar T}_f + T_{21,fG}^{\rm LF}$.

We find that the polarized contamination is significant and, in many realizations, prevents the convergence within the prior range or leads to reconstructed profiles with a high frequency trough that are, again, discarded from the analysis.
Note that we retain a reconstructed profile if these criteria are satisfied by both polarizations. In the ``all-$\phi$" case, we discard almost all realizations, concluding that the level of contamination of the simulation is too high for this scenario.
On the contrary, using the ``low $\phi$" simulations, it is possible to select a meaningful sub-sample of realizations. Indeed, in this case, we retain the reconstructed profile in both polarizations for $\sim 30\%$ of the cases (figure~\ref{fig:gauss_flat}).
As discussed in Section~\ref{sec:pol}, we also consider a more optimistic case with a magnitude of the polarized spectrum reduced to a $10\%$ value of the current simulations. Even at this reduced level of contamination, the reconstruction remains biased in a way similar to what is shown in figure~\ref{fig:gauss_flat}.

We eventually extract the $21$~cm signal using a Gaussian model $T_m = {\bar T}_f + T_{21,G}^{\rm LF}$, i.e. the same functional form used for the simulation input.
The magnitude of the polarized contamination prevents the extraction of the 21~cm signal in virtually all the simulated cases.
We find, instead, convergence for all cases when the contamination is reduced to the $10\%$ level (figure~\ref{fig:gauss_gauss}). The effect of the polarized leakage is, again, a bias similar to the one in figure~\ref{fig:flat_flat};

\item the 21~cm signal is the fiducial HF band model (section~\ref{sec:signal}). The contamination
derived from our simulations is significantly higher than the $21$~cm signal, preventing  the convergence of the extraction algorithm to a physically meaningful solution for $\Delta z$, the reionization duration. When we consider the case of a $10\%$ contamination we find that the extraction is possible, although the recovered signal
is noticeably biased (figure~\ref{fig:HF_p18}). As already noticed in section~\ref{sec:pol}, the bias is stronger in the ``low $\phi$" case (up to 10\%), where $\Delta z$ systematically tends to lower values. The bias is still present in the ``all $\phi$" case, but less pronounced.
\end{itemize}

\begin{figure}
\includegraphics[width=\columnwidth]{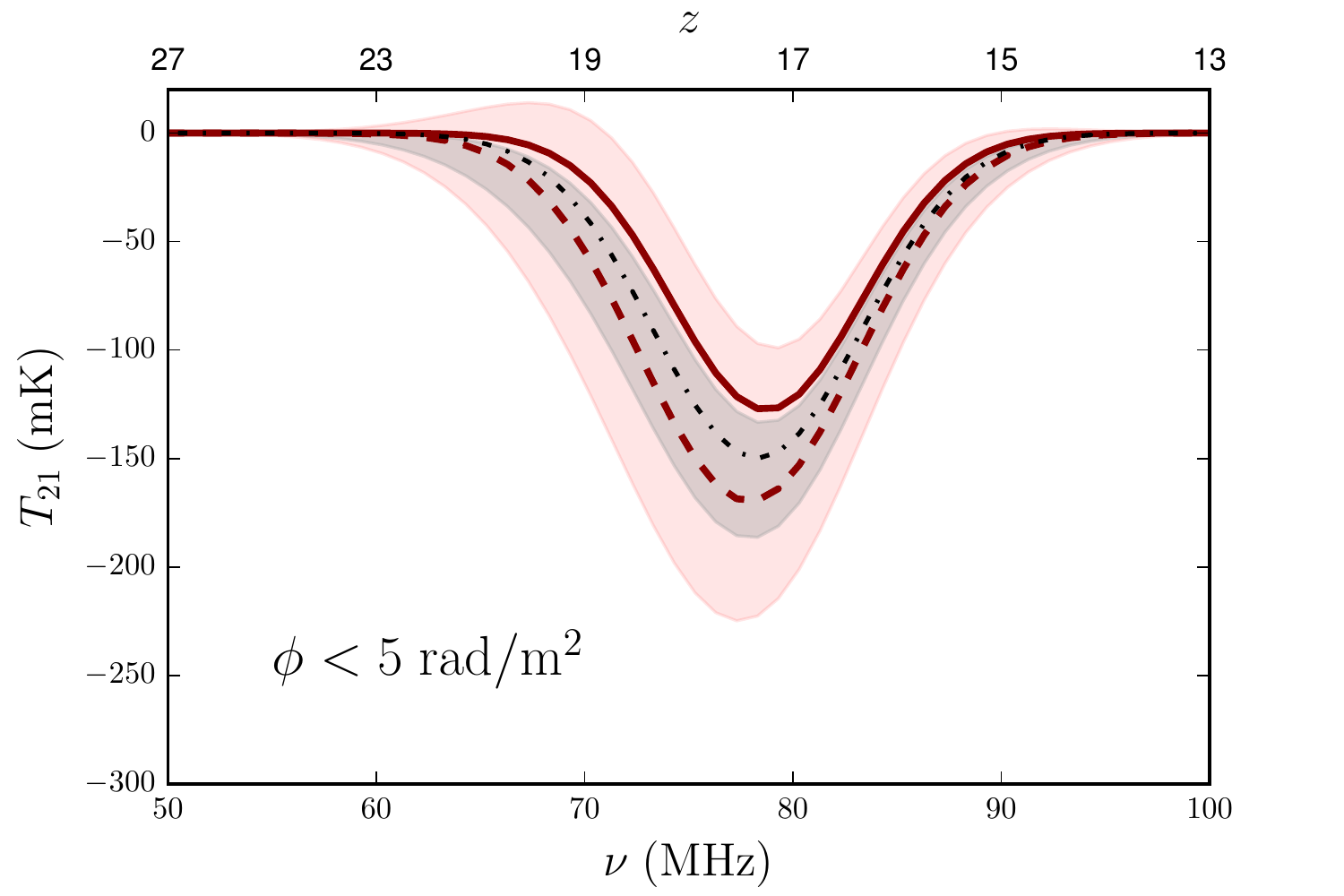}
\caption{The black dotted-dashed line is the Gaussian fiducial input signal in the LF band.
We show here reconstructed $T_{21,G}$ signal
considering the ``low $\phi$" (i.e. $\phi < 5$~rad/m$^{2}$) case with signal magnitude reduced to the $10\%$ of the reference simulation - see text for details. 
The red (grey) shaded area is the $1\sigma$ region around the mean for the $xx$ ($yy$) polarization. The solid (dashed) red line shows one of the the reconstructed $T_{21,G}$  signal for the $xx$ ($yy$) polarization. }
\label{fig:gauss_gauss}
\end{figure}

\begin{figure}
\includegraphics[width=\columnwidth]{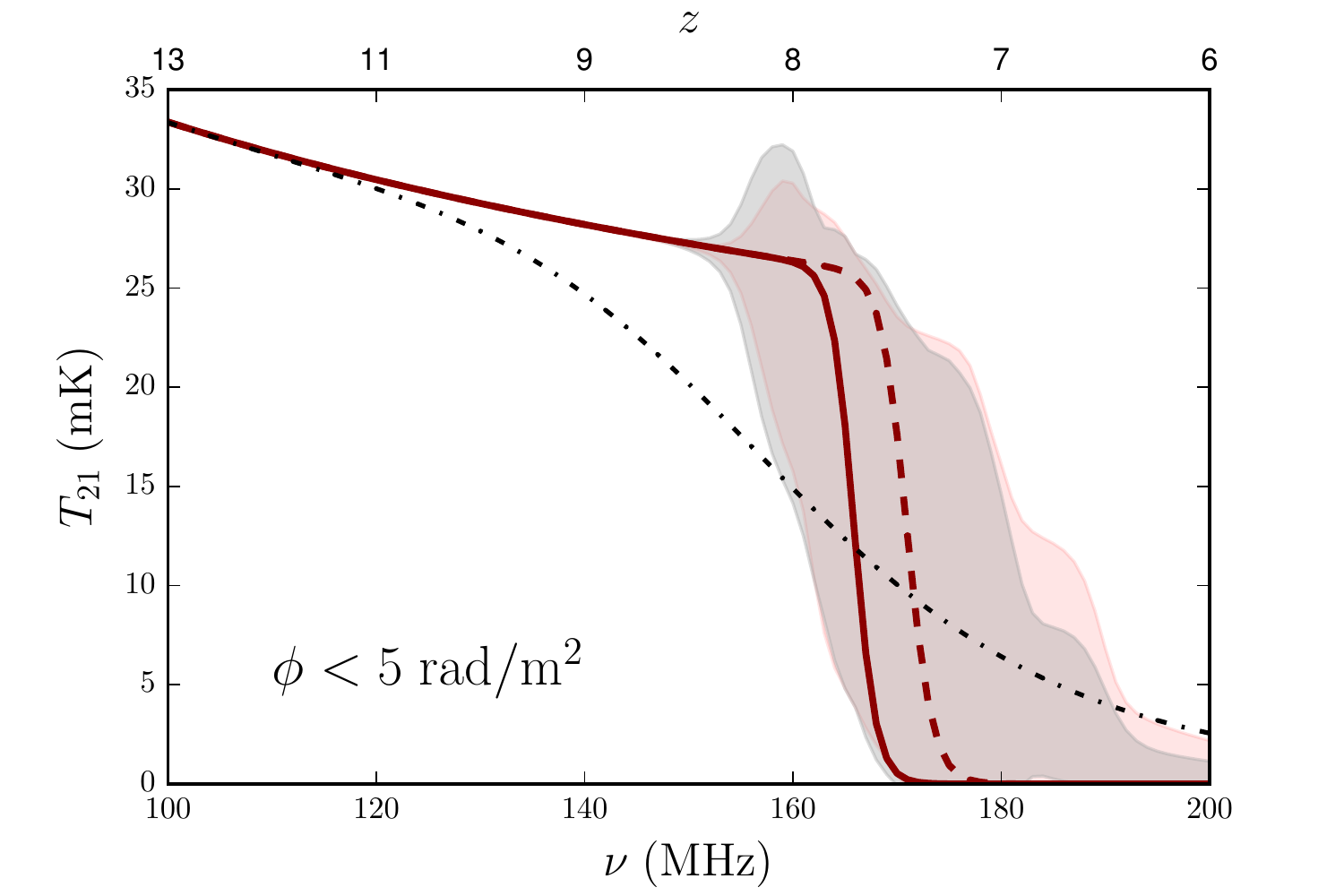}
\includegraphics[width=\columnwidth]{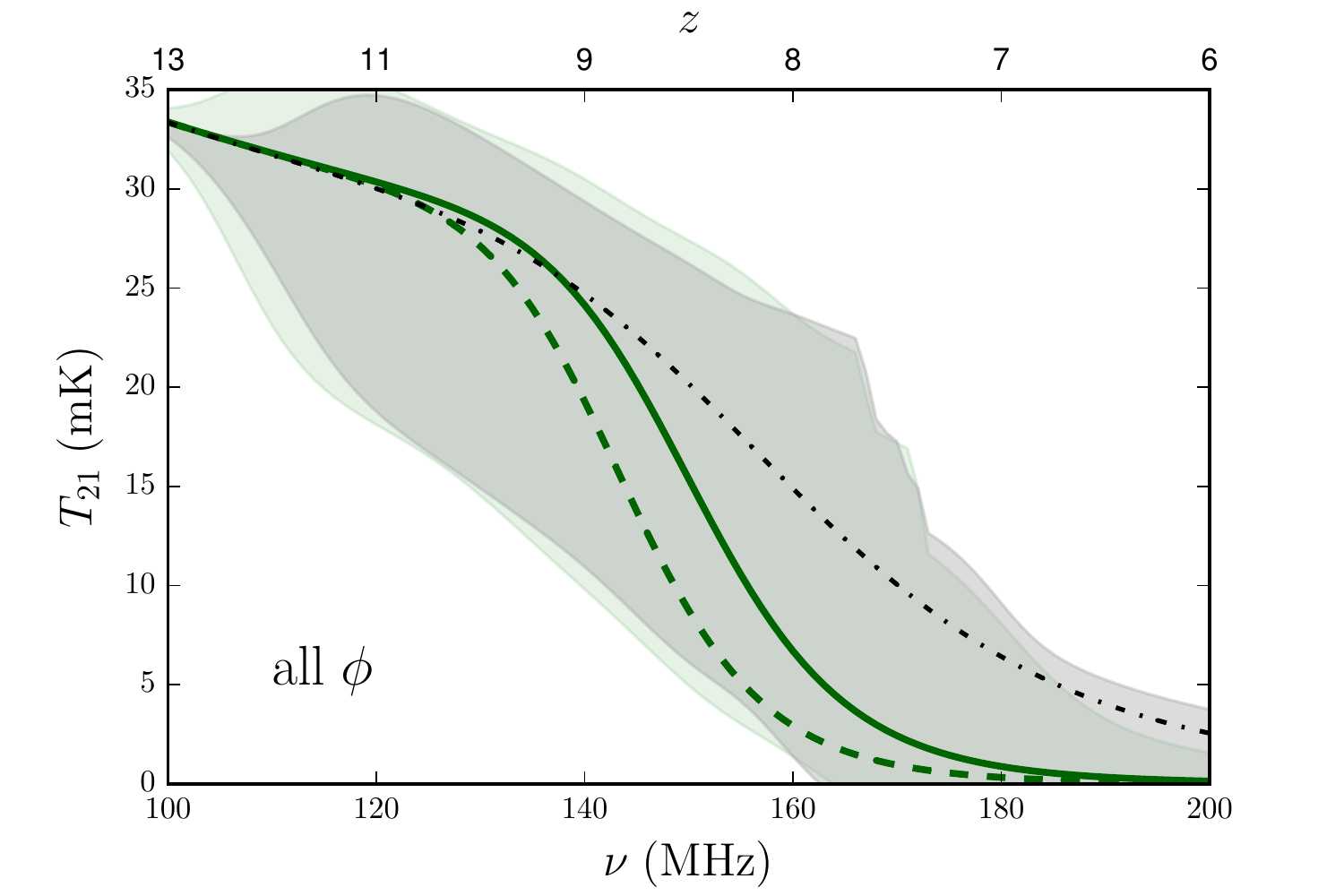}
\caption{The black dotted-dashed line in both panels is the fiducial input EoR signal. {\it Top panel}: reconstructed $T_{21}$ signal in the HF band, in the ``low $\phi$" (i.e. $\phi < 5$~rad/m$^{2}$) case with a $10\%$ reduced magnitude (see text for details).   
The solid (dashed) red line shows one of the the reconstructed $T_{21}^{HF}$  signal for the $xx$ ($yy$) polarization.  The red (grey) shaded area is the $1\sigma$ region around the mean for the $xx$ ($yy$) polarization. {\it Bottom panel}: same but for the ``all $\phi$"  simulations. The green (grey) shaded area is the $1\sigma$ region around the mean for the $xx$ ($yy$) polarization and the solid (dashed) green line shows one of the reconstructed signal for the $xx$ ($yy$) case.}\label{fig:HF_p18}
\end{figure}

\section{Discussion and Conclusions}\label{sec:con}

In this paper we have studied the impact of polarized foregrounds on the measurement of the $21$~cm global signal.
We simulated realistic observations taken with a zenith-pointing dipole, spanning an $8$~hour range with a $1$~minute cadence. Simulations include  the all-sky polarized foreground template maps from \citet{PolSynch} and a realistic dipole beam in order to generate polarized spectra. We also include a different polarized template where the contamination is reduced to low Faraday depth values, i.e. $\phi < 5$~rad/m$^{2}$. We simulate two antenna orientations ($xx$ and $yy$) separately, using the corresponding beam models. We also consider a more optimistic case where the amplitude is $10\%$ of the template maps in order to empirically account for depolarization effects not included in the \citet{PolSynch} model. Total intensity foregrounds are directly modelled through their spectra, as a 4$^{\rm th}$-order log-polynomial function. 

We included three different 21~cm global signal models: a fiducial EoR $\tanh$ model in the $100-200$~MHz (HF) range, a fiducial Gaussian and a flattened Gaussian \citep{EDGES} absorption profile in the $50-100$~MHz range (LF).
We performed a Bayesian extraction of the global $21$~cm signal from the simulated spectra. 

We draw a few main conclusions from our work. 
We find that, generally, the contamination from our polarized foreground model has a magnitude and frequency behaviour that prevents the extraction of the 21~cm fiducial signal 
both in the HF and LF bands. In order to detect the signal, the contamination needs to be fainter: at the $\sim 10\%$ magnitude level, the extraction of the $21$~cm signal in both bands is possible, but is significantly biased. In the HF band, the middle point of reionization is biased up to the $10\%$ level and the duration of reionization is poorly recovered, underestimated by a factor up to $10$. In the LF band, the bias affects the amplitude of the fiducial Cosmic Dawn Gaussian signal at the $20\%$ level.

The contamination from polarization leakage can be mitigated by the subtraction of the two orthogonal polarizations observed by a dual polarization antenna. Asymmetries in the beam pattern as well as errors in the relative calibration of the two polarizations can still, however, introduce  polarization contamination at some level. By reducing the magnitude of the polarized signal to the $10\%$ level, we mimic this case too and show that the contamination may not be negligible even in dual polarization observations, in particular for the fiducial EoR model. For example, \citet{Monsalve2017} find a periodic residual signal at the 30~mK level that could be consistent with polarization contamination.

In the light of the detection of the Cosmic Dawn signal reported by \citet{EDGES}, we include their flattened Gaussian absorption model in our simulations. We test a case where the simulation input is the fiducial Cosmic Dawn Gaussian absorption that we, however, model as a flattened Gaussian profile in the extraction. We find that in this case the signal extraction is possible even at the level of polarized intensity predicted by our simulations, if we consider the ``low $\phi$" (i.e. $\phi < 5$~rad/m$^{2}$) realizations. 
We find that the polarization
contamination tends to introduce a bias in the recovered $21$~cm signal, increasing both its amplitude and width for both polarization orientations, leading to a profile similar to what \citet{EDGES} observed. Due to the modeling uncertainties, the bias evidence remains statistically weak, i.e. in tension with the input fiducial Gaussian signal only at the $\sim 1.5 \sigma$ level.

In order to exclude the contamination from polarized foregrounds, \citet{EDGES} carried out two measurements where the dipole antenna was rotated by $90^\circ$. The best fit signal was consistent in both cases, with a $10-20\%$ difference in amplitude \citep[see Figure~2 in][]{EDGES}. We find that the difference between the $21$~cm signal extracted from $xx$ and $yy$ polarization orientations is at a similar level in our simulated cases. This result indicates that measurements with a rotated antenna do not necessarily exclude the polarized contamination and implies that the use of a dual  polarization antenna would not automatically remove the problem of polarized foregrounds.

We also simulate the case with a flattened Gaussian profile as both simulation input and model in the extraction. We find that the signal extraction is possible in the $80\%$ of runs for both the ``all $\phi$" and the ``low $\phi$" (i.e. $\phi < 5$~rad/m$^{2}$) cases, as the input $21$~cm signal is brighter than the polarized foreground. The amplitude of the extracted $21$~cm profile, however, has an amplitude bias at the $\sim 20-30\%$ level that changes with the polarization orientation which is, again, qualitatively comparable with the difference shown by \citet{EDGES} when the two polarizations are rotated by $90^\circ$. A polarized contamination, enhancing the reconstructed signal, could mitigate the need to explain the anomalously high amplitude in term of exotic physics.

\section*{acknowledgements}
MS thanks Junaid Townsend for providing the {\it SimFast21} outputs.
MS and MGS are supported by the South African Square Kilometre Array Project and National Research Foundation. MS and GB acknowledge funding from the INAF PRIN-SKA 2017 project 1.05.01.88.04 (FORECaST). We acknowledge the support from the Ministero degli Affari Esteri della Cooperazione Internazionale - Direzione Generale per la Promozione del Sistema Paese Progetto di Grande Rilevanza ZA18GR02 and the National Research Foundation of South Africa (Grant Number 113121) as part of the ISARP RADIOSKY2020 Joint Research Scheme. GB acknowledges the Rhodes University research office and support from the Royal Society and the Newton Fund under grant NA150184. This work is based on research supported in part by the National Research Foundation of South Africa (grant No. 103424).

\bibliographystyle{mnras}
\bibliography{21global_pol}

\bsp	
\label{lastpage}
\end{document}